\documentclass[utf8]{FrontiersinHarvard} 

\usepackage{url,hyperref,lineno,microtype,subcaption}
\usepackage[onehalfspacing]{setspace}
\usepackage{longtable}
\def\keyFont{\fontsize{8}{11}\helveticabold }
\def\firstAuthorLast{Payne et al.}
\def\Authors{Chandler C. Payne\,$^{1}$,Kai A. Hackney\,$^{1,}$, Lucas Guarenti Zangari\,$^{1}$, Sterling R. Kalogeras\,$^{1}$, Emmanuel Munoz\,$^{1}$, Juan Sebastián Sánchez-Gómez\,$^{2}$, Olufisayo Omojokun\,$^{1}$, and Pedro Guillermo Feijóo-García\,$^{1,*}$}
\def\Address{$^{1}$College of Computing, Georgia Institute of Technology, Atlanta, GA, USA\\$^{2}$Facultad de Ingeniería, Universidad El Bosque, Bogotá D.C., Colombia}
\def\corrAuthor{Pedro Guillermo Feijóo-García}
\def\corrEmail{pfeijoogarcia@gatech.edu}

\begin{document}
\onecolumn
\firstpage{1}

\title[Exploring Community Perceptions of Academic Dishonesty]{\textit{Same Rules, Mixed Messages:} Exploring Community Perceptions of Academic Dishonesty in Computing Education} 

\author[\firstAuthorLast ]{\Authors} 
\address{} 
\correspondance{} 

\extraAuth{}

\maketitle

\begin{abstract}
Academic dishonesty has long been a concern in computing education, and the rapid growth of online learning and generative artificial intelligence (AI) has further complicated how cheating is perceived and addressed. We report on a study examining how different actors in the computer science (CS) classroom interpret potential cheating scenarios and the motivations behind academic dishonesty. Participants included instructors (n = 6), teaching assistants (TAs; n = 22), and undergraduate students (n = 538) enrolled in two CS courses at a large Southeastern institution in the United States. Respondents classified scenarios as serious cheating, trivial cheating, or not cheating and responded to an open-ended question about motivations for academic dishonesty. Our findings reveal differences in perceptions across the three stakeholder groups. Among the instructors who participated, grade pressure and laziness were the most frequently mentioned motivations for academic dishonesty, whereas students and TAs more frequently emphasized gaps in prerequisite knowledge and time management challenges. Given the small instructor sample, these qualitative findings should be interpreted as descriptive and exploratory rather than representative of instructors more broadly. These results highlight differences in how academic dishonesty is understood across classroom stakeholders and underscore the need for clearer communication of academic-integrity expectations and instructional strategies that help students interpret and navigate those expectations in computing education, particularly in post-COVID learning environments where hybrid instruction, increased reliance on digital resources, and AI-assisted tools have reshaped students' approaches to coursework and learning.

\tiny
 \keyFont{ \section{Keywords:} Student Integrity, Computing Education, Community Perceptions, AI in Education, Higher Education} 
\end{abstract}

\section{Introduction}
Academic integrity has long been a central concern in computing education. Programming assignments, online assessments, and widespread access to digital resources create environments where collaboration and external assistance can easily blur into academic misconduct \cite{5,6}. In recent years, these challenges have intensified with the emergence of generative artificial intelligence (GenAI) systems capable of producing functional code and solving programming problems \cite{14,15}. At the same time, the rapid expansion of online and hybrid instruction following the COVID-19 pandemic has further complicated how students interact with peers, resources, and assessment materials. Together, these developments raise an important question for computing eduction: not solely how cheating occurs, but how different actors in the classroom interpret what counts as cheating in the first place.

Maintaining academic integrity is a priority in higher education, both to ensure that students achieve the intended learning outcomes and to preserve the credibility of academic credentials \cite{6}. In computing classrooms, this issue can be particularly complex. Collaboration, reuse of code, and consulting external resources are often legitimate and even encouraged as part of authentic professional practice. However, these same practices can also lead to ambiguity when students attempt to determine what constitutes acceptable collaboration versus academic misconduct. Previous studies have reported on miscommunication and how unclear definitions of cheating can lead to differing perceptions among students and instructors regarding what behaviors constitute academic dishonesty \cite{3}. As a result, instructors, teaching assistants, and students may hold differing interpretations of where these boundaries lie. Yet relatively less attention has been given to examining these interpretations across multiple classroom stakeholder groups using the same concrete academic-integrity scenarios, particularly in contemporary computing courses shaped by online learning and GenAI.

The computing education community has made significant efforts to understand and mitigate academic dishonesty. Prior work has examined why students cheat and explored strategies to discourage or detect such behaviors. These efforts include algorithmic detection approaches such as rare-answer analysis \cite{1}, style anomalies in code submissions \cite{8}, and temporal analysis of student activity during assessments \cite{11,4}. Other work has explored structural and pedagogical interventions, including the use of automated problem generators to reduce answer sharing \cite{12}, watermarked responses to detect collaboration \cite{2}, and instructional practices that promote academic integrity through clearer communication, integrity discussions, and supportive classroom environments \cite{9,10,13}. Together, these approaches demonstrate the range of technical and pedagogical strategies developed to address academic misconduct in computing education. However, approaches designed to detect, prevent, or discourage misconduct do not necessarily explain whether students and instructional staff interpret potentially dishonest behaviors in the same way. Understanding these differences is important because policies and interventions ultimately depend on shared interpretations of what behaviors are considered acceptable or unacceptable.

However, despite these efforts, academic integrity remains a persistent and evolving challenge. The introduction of GenAI tools such as ChatGPT has created new opportunities for students to outsource work in ways that are difficult to detect using traditional approaches \cite{14}. Research has shown that these systems are capable of producing reasonable solutions to introductory programming tasks and can iteratively refine their outputs, leaving many existing forms of assessment vulnerable to solutions generated with artificial intelligence (AI) \cite{15}. At the same time, prior research suggests that even before the rise of GenAI, ambiguity around collaboration and academic integrity policies already contributed to differing interpretations of cheating within computing courses \cite{3,13}. These developments further highlight the importance of understanding how academic integrity is interpreted by different actors in the classroom: students, teaching assistants, and instructors. In particular, prior work leaves unresolved how these stakeholder groups evaluate the same potentially ambiguous situations and whether they identify similar reasons that may lead students toward academic dishonesty.

To address this gap, we investigate how students, instructors, and teaching assistants interpret and reason about scenarios involving potential academic dishonesty in computer science (CS) courses. Our work builds on the early-2000s study by \cite{3} by examining how these groups respond to situations that may constitute cheating and the reasons they believe may lead individuals toward academic misconduct. Specifically, we revisit and extend this scenario-based line of inquiry by examining students, teaching assistants, and instructors together, adapting the investigation to contemporary computing-education contexts, and considering academic-integrity situations shaped by the widespread availability of GenAI tools. In doing so, we revisit this earlier line of inquiry within the substantially changed context of post-COVID computing education. 

By analyzing these perspectives together, this study aims to provide insights into how differing interpretations of academic integrity and the perceived motivations behind academic dishonesty may shape decision-making and behaviors in the CS classroom. The following research questions guide our study:

\begin{enumerate}
    \item \textbf{RQ1}:  How do students, instructors, and teaching assistants respond to scenarios suggesting academic dishonesty in the CS classroom?
    \item \textbf{RQ2}: What reasons do students, instructors, and teaching assistants in CS consider that could lead someone toward academic dishonesty in the CS classroom?
\end{enumerate}

An earlier version of this work appeared as a conference abstract at the 21st ACM Conference on International Computing Education Research (ICER `25) \cite{payne2025exploring}. The present article substantially extends that preliminary report with additional analysis, expanded results, and a deeper discussion of our findings. 

\section{Related Work}
Cheating remains a persistent concern in the classroom, especially in computer science classes and online assessments \cite{5}. The emergence of generative artificial intelligence (GenAI) has further complicated this challenge because some traditional detection approaches may be less effective when students use systems such as ChatGPT to generate or substantially complete academic work \cite{14}. \cite{15} demonstrated that ChatGPT is capable of generating reasonable responses to introductory Python programming tasks and iteratively refining its own work, which leaves traditional assessments vulnerable to students outsourcing portions of their work to AI systems. Together, these developments show how evolving technologies continue to challenge existing approaches to academic-integrity enforcement and create a need to better understand how cheating is both detected and perceived.

One prominent line of research has focused on detecting academic misconduct through patterns in student submissions and assessment behavior. There are many methods by which to detect cheating, but several rely on different forms of evidence. Rare answers represent one such approach; \cite{1} define them as answers, correct or incorrect, that are shared by a very small portion of the class, with what qualifies as rare depending in part on class size. \cite{8} expand on this further with the concept of style anomalies, which aims to determine if students are using unauthorized outside resources through similarity checks, since those resources may provide code in a different manner than what is taught in the respective class. Along with rare answers, \cite{11} used temporal evidence to determine how long students were reading questions, or whether patterns suggested that answers were being copied and pasted without meaningful engagement with the question. Temporal analysis can also be used to detect cheating by examining when students answer individual questions and submit an assessment, as seen through \cite{4}. Although these approaches provide useful ways to identify potentially suspicious behavior, they primarily focus on detecting misconduct rather than understanding whether students and instructional staff agree on what should be considered misconduct in the first place.

A second body of work has examined structural and instructional approaches intended to reduce opportunities or motivations for academic dishonesty. Cheating enforcement can impose substantial time demands on instructors and require significant institutional resources \cite{6}. Albluwi \cite{7} explains how open repositories, reused assignments, and a lack of plagiarism checks can offer easy opportunities for students to commit academic misconduct. Because continuously monitoring and redesigning assessments can require considerable instructor effort, researchers have also explored lower-effort interventions. \cite{10} explore how practices such as normalizing help-seeking, holding integrity discussions, explaining cheating-detection tools, and using integrity quizzes may reduce indicators of academic dishonesty.

Automatic approaches can also reduce some of the instructional workload associated with preventing and identifying misconduct. \cite{12} offer an automatic problem generator that both creates and grades assignments, making it feasible to provide different versions of an assessment to individual students without placing the full burden on instructors and teaching assistants. This approach can also make direct answer sharing less useful when students receive different problem instances. \cite{2} specifically address collaborative answer sharing through the use of a watermarked answer. This provides another way to detect collaboration when students copy and paste answers from classmates, without relying solely on rare answers or temporal evidence. Taken together, these structural, pedagogical, and automated approaches seek to reduce either the opportunity or the motivation for academic misconduct. However, like detection-oriented approaches, they do not necessarily tell us whether students, teaching assistants, and instructors interpret the boundaries of acceptable behavior in the same way.

A complementary line of research focuses more directly on communication, expectations, and perceptions of academic integrity. Studies show that miscommunication and a lack of direct definitions of cheating can lead to differing perceptions of what qualifies as cheating \cite{3}. \cite{9} show how professor-student relationships and a proper support system can improve that communication and decrease students' motivation to engage in academic dishonesty. \cite{13} further emphasize that clear expectations, explicit instructions, and well-defined boundaries can decrease occurrences of dishonesty. This work suggests that academic integrity is not solely a problem of detecting or preventing prohibited behavior. It is also a problem of making sure that students and instructional staff understand the boundaries between acceptable assistance, collaboration, and resource use and academic misconduct. However, comparatively less is known about how students, teaching assistants, and instructors respond to the same potentially ambiguous situations in contemporary computing classrooms. This unresolved question motivates our examination of academic-integrity perceptions across these three stakeholder groups.

\section{Method}

We conducted an online, asynchronous study during the Fall 2024 semester, recruiting participants from the Georgia Institute of Technology--a large Southeastern institution in the United States. The sample included undergraduate students enrolled in computing courses, teaching assistants (TAs) supporting those courses, and faculty members from the institute's College of Computing. Students were recruited from two large-enrollment courses in the College of Computing. The first was the second programming course offered in the undergraduate curriculum, focused on object-oriented programming. The second was an introductory software design and engineering course that has the programming course as a prerequisite. Teaching assistants were undergraduate TAs supporting these courses, while faculty participants were also recruited from the College of Computing.

Data were collected via a structured questionnaire, which was distributed by instructors through the university's Learning Management System (LMS). Participation was voluntary. Students received extra credit for participating in the study, while no incentives were offered to teaching assistants or faculty members. Participation required approximately 15 minutes, and all communication was handled asynchronously via email. We were not able to reliably determine the total number of individuals who received the invitation to participate and, therefore, do not report response rates. The study received approval from the Institutional Review Board (IRB) prior to data collection.

\subsection{Participants}
Our study included a total of n = 566 participants, distributed across three core groups: faculty (n = 6), teaching assistants (TAs; n = 22), and students (n = 538). Because the number of participants differed substantially across these groups, particularly for faculty, comparisons across stakeholder groups should be interpreted with caution. Findings from the faculty group are descriptive of the six instructors who participated and are not intended to represent computing faculty more broadly.

The student sample represented a diverse set of demographic backgrounds: n = 375 identified as male, n = 149 as female, and n = 4 as non-binary. Ten students chose not to disclose their gender. In terms of ethnicity, most students identified as Asian (n = 348) or White (n = 137; non-Hispanic/non-Latin American). Additional identities included Hispanic/Latin American (n = 33), Black/African American (n = 38), Middle Eastern/North African (n = 18), and Native American/Alaska Native (n = 3).

All student participants were undergraduates, with the majority identifying as domestic students (n = 475) and a smaller portion as international students (n = 63). Participants were all older than 18 years, with most aged 22 or younger (n = 535); only three students (n = 3) fell between 25 and 29 years of age. On average, students reported taking 14.8 credit hours (SD = 2.1). Most were in their first year (n = 259), followed by second-year students (n = 208), and n = 71 students were in their third year or beyond. Additionally, most students (n = 378) reported no prior industry job or internship experience related to computing.

Among the teaching assistants (n = 22), the majority were male (n = 16), followed by female (n = 5), and one identified with another gender. All teaching assistants were older than 18 years of age, and none exceeded 22 years. The predominant ethnic groups were Asian (n = 11) and White (n = 10; non-Hispanic/non-Latin American), with two TAs selecting other ethnic identities. Most TAs were domestic (n = 20), and the sample was evenly split between second-year students (n = 11) and those in their third year or beyond (n = 11).

Finally, the faculty group (n = 6) consisted entirely of domestic participants, primarily male (n = 4), with one female and one identifying with another gender group. Ethnic representation included one (n = 1) Asian, two (n = 2) Black/African American, one (n = 1) Hispanic, and one (n = 1) White (non-Hispanic/non-Latin American) participant. One (n = 1) faculty member did not disclose their ethnicity. Participants had a wide range of institutional experience, from two to 35 years. As noted above, findings involving faculty should be interpreted as descriptive given the small number of faculty participants.

\subsection{Procedure}\label{procedure}

Data collection took place during the Fall 2024 academic semester via an online and asynchronous questionnaire. Participants first reviewed an informed consent statement and indicated their consent to participate and then proceeded to the main set of questions. First, students, faculty, and teaching assistants (TAs) were asked to classify 13 scenarios using three options: 1) Not Cheating, 2) Trivial Cheating, and 3) Serious Cheating. The first ten scenarios were adapted from the work by \cite{3} to better reflect the context and practices of the CS courses included in our study. We maintained the situations and ideas represented in the original scenarios while updating contextual details that were no longer applicable to our courses. For example, because neither course uses computer labs, a scenario originally framed as "You are working in a computer lab" was adapted to "You are working on a programming assignment." We also added three scenarios (S11-S13) to reflect practices that have become increasingly relevant in contemporary computing education, particularly the use of GenAI tools and online-based course formats. These scenarios were discussed and tailored within the research team, following the context and structure of the scenarios presented in the original study. The 13 scenarios were as follows:

\begin{itemize}
\item \textbf{S1:} You are working on a programming assignment. A classmate is having difficulty with their program. You look at their code to help identify the error.

\item \textbf{S2:} You are working on a programming assignment. A classmate is having difficulty with their program. You show the classmate a similar section of your code to help them understand.

\item \textbf{S3:} You are given an example of an already-compiled program that is executable. Your assignment is to create a program that runs like the example program. You decompile the example program and use parts of the resulting code in your assignment.

\item \textbf{S4:} At a review session, the TA goes over the types of questions you need to study for the exam. The TA gives example questions and answers them. When you receive the exam, you realize that the TA gave you the exact questions for the exam. Consider the TA's actions.

\item \textbf{S5:} You have spent three hours working on a portion of your assignment and you are having difficulty understanding it. There are resources (notes and/or assignments) from previous terms that answer your question. You look at those resources long enough to gain understanding. You have learned from those resources. You now use the information from those resources to finish your assignment.

\item \textbf{S6:} You use resources (notes and/or assignments) from a previous semester while studying for a quiz. When you take the quiz, it is identical to the resources that you found. You repeat all of the answers from the resources verbatim. Some of the answers are essay (open-ended) questions.

\item \textbf{S7:} You have an assignment due. However, you have not yet had time to complete it due to an overload of coursework. You get a time exemption by telling your professor that you have been sick.

\item \textbf{S8:} Your syllabus lists a website that you are allowed to use. You use an algorithm from this website without citing the source.

\item \textbf{S9:} In recitation one week, your TA goes over the type of questions you need to study for next week's quiz. Your TA gives example questions and then answers them. When you get the quiz the next week you realize your TA gave the exact questions from the quiz. You write the exact answers you were given in recitation.

\item \textbf{S10:} Your professor has forbidden group work on a particularly difficult homework assignment. You work on the assignment with someone else from the class.

\item \textbf{S11:} A student writes their own code for an assignment but runs into an error. They provide the code to ChatGPT and ask for help debugging the code.

\item \textbf{S12:} You are in an online version of the class you are taking now. You have an open-note exam and use the internet during said exam.

\item \textbf{S13:} Trying to better understand the material, a student asks ChatGPT to explain a concept. In an effort to help visualize the learning process, ChatGPT provides a section of code explaining the concept, and the student then understands where they went wrong. They use a small portion of code from the provided ChatGPT answer to edit the code they were having trouble with.

\end{itemize}

Additionally, one open-ended question asked participants to share what reasons they believed could lead someone to cheat in their CS class.

Following these questions, participants were asked to respond to a set of demographic questions specific to each group: 14 for students, nine for TAs, and six for instructors. For the scope of this paper, we report on all participants' gender, ethnicity, and domestic status, as well as the following demographics:

\begin{itemize}
\item \textbf{Students'} academic year, prior work experience, and self-reported religiosity.
\item \textbf{TAs'} academic year.
\item \textbf{Faculty} institutional experience (i.e., years at the institution).
\end{itemize}

Additionally, we measured participants' self-reported religiosity by asking how much they agreed with the statement, \textit{``I am a religious person.''} Responses were provided on a seven-point scale ranging from 1 (Totally disagree) to 7 (Totally agree), with higher values indicating greater self-reported religiosity. Because this measure was ordinal, we used Spearman's rank-order correlation when examining its association with responses to the academic-integrity scenarios.

\section{Data Analysis}\label{analysis}
Our analysis followed a mixed-methods approach, combining both quantitative and qualitative strategies. For the close-ended responses, we used descriptive statistics to examine how students, TAs, and instructors classified each of the 13 scenarios. Given the substantially different sample sizes across the three participant groups, particularly the small faculty sample, comparisons across groups were interpreted primarily as descriptive.

Within the student population, we further examined associations between selected demographic characteristics and responses to the academic-integrity scenarios. We used Spearman's rank-order correlation to examine associations involving age and self-reported religiosity. Because these analyses involved multiple comparisons across the 13 scenarios, we applied the Holm-Bonferroni correction to account for the increased likelihood of Type I error. In interpreting these results, we considered both statistical significance and the magnitude of the observed correlations.

For the open-ended responses, we conducted a thematic analysis following the process described by \cite{thematic} to identify recurring categories (i.e., themes) across participants' responses. The first author conducted the coding under the supervision of the corresponding author. The process began with a review of the complete set of responses to become familiar with the data, followed by coding the responses according to the ideas represented in them. These codes were then organized into candidate themes, which were iteratively reviewed and refined to ensure that the categories appropriately represented the responses assigned to them. Theme definitions and boundaries were revisited throughout this process before the final categories were established. Each response was ultimately assigned to the theme that most closely represented its primary explanation for academic dishonesty. Consequently, the percentages reported for each theme represent the proportion of participants within each stakeholder group whose response was assigned to that theme. Because the coding was conducted by one researcher, we did not perform an inter-rater reliability or coder-agreement assessment.

\section{Findings on Participants' Perceptions}\label{analysis}

For the close-ended responses, we employed both descriptive and inferential statistics to examine group differences and response distributions. Our quantitative analysis builds on the framework established by \cite{3}, organizing results by the three primary participant groups: faculty (n = 6), teaching assistants (TAs; n = 22), and students (n = 538). Given the substantially smaller faculty sample, findings involving instructors are presented descriptively and should be interpreted with caution.

Further exploring these responses through the demographic data we collected, within the student population we initially observed statistically significant negative correlations between age and responses to scenarios S4 ($r_s = -.135$), S8 ($r_s = -.136$), S10 ($r_s = -.088$), S12 ($r_s = -.085$), and S13 ($r_s = -.104$). We also observed a statistically significant negative correlation between self-reported religiosity and responses to S4 ($r_s = -.124$). These correlations were small in magnitude, indicating limited associations between these characteristics and participants' classifications of the scenarios.

After applying the Holm-Bonferroni correction for multiple comparisons, the associations between age and S4 and S8, as well as the association between self-reported religiosity and S4, remained statistically significant. The initially observed associations between age and S10, S12, and S13 did not remain significant after correction and are therefore treated only as descriptive patterns. Overall, higher age and greater self-reported religiosity were associated with slightly lower cheating classifications in the scenarios where corrected significance was retained, although the magnitude of these associations remained small. We found limited or no statistically significant relationships in the scenarios with regard to race, gender, domestic/international status, and work experience. Table \ref{tableResults} contains the full data for the percentages related to each scenario and actor group within the classroom setting.

\begin{table}[!ht]
\centering
\caption{\centering\label{tableResults}Response Frequency per Scenario and Participant Group} 
\resizebox{\textwidth}{!}{%
\begin{tabular}{|c|ccc|ccc|ccc|}
\hline
\multicolumn{1}{|l|}{} & \multicolumn{3}{c|}{Not Cheating} & \multicolumn{3}{c|}{Trivial Cheating} & \multicolumn{3}{c|}{Serious Cheating} \\ \hline
Scenario & \multicolumn{1}{c|}{\begin{tabular}[c]{@{}c@{}}Faculty \\      (n = 6)\end{tabular}} & \multicolumn{1}{c|}{\begin{tabular}[c]{@{}c@{}}Teaching Assistants\\      (n = 22)\end{tabular}} & \begin{tabular}[c]{@{}c@{}}Students\\      (n = 538)\end{tabular} & \multicolumn{1}{c|}{\begin{tabular}[c]{@{}c@{}}Faculty \\      (n = 6)\end{tabular}} & \multicolumn{1}{c|}{\begin{tabular}[c]{@{}c@{}}Teaching Assistant\\      (n = 22)\end{tabular}} & \begin{tabular}[c]{@{}c@{}}Student\\      (n = 538)\end{tabular} & \multicolumn{1}{c|}{\begin{tabular}[c]{@{}c@{}}Faculty \\      (n = 6)\end{tabular}} & \multicolumn{1}{c|}{\begin{tabular}[c]{@{}c@{}}Teaching Assistants\\      (n = 22)\end{tabular}} & \begin{tabular}[c]{@{}c@{}}Students\\      (n = 538)\end{tabular} \\ \hline
S1 & \multicolumn{1}{c|}{83.3\%} & \multicolumn{1}{c|}{86.4\%} & 76.0\% & \multicolumn{1}{c|}{0.0\%} & \multicolumn{1}{c|}{13.6\%} & 22.5\% & \multicolumn{1}{c|}{16.7\%} & \multicolumn{1}{c|}{0.0\%} & 1.5\% \\ \hline
S2 & \multicolumn{1}{c|}{16.7\%} & \multicolumn{1}{c|}{31.8\%} & 37.7\% & \multicolumn{1}{c|}{50.0\%} & \multicolumn{1}{c|}{59.1\%} & 54.5\% & \multicolumn{1}{c|}{33.3\%} & \multicolumn{1}{c|}{9.1\%} & 7.8\% \\ \hline
S3 & \multicolumn{1}{c|}{0.0\%} & \multicolumn{1}{c|}{50.0\%} & 29.0\% & \multicolumn{1}{c|}{16.7\%} & \multicolumn{1}{c|}{36.4\%} & 41.6\% & \multicolumn{1}{c|}{83.3\%} & \multicolumn{1}{c|}{13.6\%} & 29.4\% \\ \hline
S4 & \multicolumn{1}{c|}{50.0\%} & \multicolumn{1}{c|}{36.4\%} & 30.5\% & \multicolumn{1}{c|}{0.0\%} & \multicolumn{1}{c|}{22.7\%} & 22.5\% & \multicolumn{1}{c|}{50.0\%} & \multicolumn{1}{c|}{40.9\%} & 47.0\% \\ \hline
S5 & \multicolumn{1}{c|}{66.7\%} & \multicolumn{1}{c|}{86.4\%} & 89.4\% & \multicolumn{1}{c|}{33.3\%} & \multicolumn{1}{c|}{13.6\%} & 9.1\% & \multicolumn{1}{c|}{0.0\%} & \multicolumn{1}{c|}{0.0\%} & 1.5\% \\ \hline
S6 & \multicolumn{1}{c|}{50.0\%} & \multicolumn{1}{c|}{18.2\%} & 30.9\% & \multicolumn{1}{c|}{0.0\%} & \multicolumn{1}{c|}{36.4\%} & 36.6\% & \multicolumn{1}{c|}{50.0\%} & \multicolumn{1}{c|}{45.5\%} & 32.5\% \\ \hline
S7 & \multicolumn{1}{c|}{0.0\%} & \multicolumn{1}{c|}{4.5\%} & 18.0\% & \multicolumn{1}{c|}{50.0\%} & \multicolumn{1}{c|}{50.0\%} & 56.3\% & \multicolumn{1}{c|}{50.0\%} & \multicolumn{1}{c|}{45.5\%} & 25.7\% \\ \hline
S8 & \multicolumn{1}{c|}{66.7\%} & \multicolumn{1}{c|}{50.0\%} & 27.9\% & \multicolumn{1}{c|}{16.7\%} & \multicolumn{1}{c|}{31.8\%} & 50.7\% & \multicolumn{1}{c|}{16.7\%} & \multicolumn{1}{c|}{18.2\%} & 21.4\% \\ \hline
S9 & \multicolumn{1}{c|}{83.3\%} & \multicolumn{1}{c|}{63.6\%} & 48.0\% & \multicolumn{1}{c|}{0.0\%} & \multicolumn{1}{c|}{22.7\%} & 27.3\% & \multicolumn{1}{c|}{16.7\%} & \multicolumn{1}{c|}{13.6\%} & 24.7\% \\ \hline
S10 & \multicolumn{1}{c|}{0.0\%} & \multicolumn{1}{c|}{0.0\%} & 3.0\% & \multicolumn{1}{c|}{16.7\%} & \multicolumn{1}{c|}{68.2\%} & 41.6\% & \multicolumn{1}{c|}{83.3\%} & \multicolumn{1}{c|}{31.8\%} & 55.4\% \\ \hline
S11 & \multicolumn{1}{c|}{33.3\%} & \multicolumn{1}{c|}{59.1\%} & 33.6\% & \multicolumn{1}{c|}{16.7\%} & \multicolumn{1}{c|}{36.4\%} & 47.4\% & \multicolumn{1}{c|}{50.0\%} & \multicolumn{1}{c|}{4.5\%} & 19.0\% \\ \hline
S12 & \multicolumn{1}{c|}{16.7\%} & \multicolumn{1}{c|}{40.9\%} & 20.4\% & \multicolumn{1}{c|}{0.0\%} & \multicolumn{1}{c|}{13.6\%} & 24.7\% & \multicolumn{1}{c|}{83.3\%} & \multicolumn{1}{c|}{45.5\%} & 54.8\% \\ \hline
S13 & \multicolumn{1}{c|}{16.7\%} & \multicolumn{1}{c|}{59.1\%} & 43.7\% & \multicolumn{1}{c|}{33.3\%} & \multicolumn{1}{c|}{36.4\%} & 48.3\% & \multicolumn{1}{c|}{50.0\%} & \multicolumn{1}{c|}{4.5\%} & 8.0\% \\ \hline
\end{tabular}%
}
\end{table}

\begin{table}[t]
\centering
\caption{Reported Reasons for Academic Dishonesty Across Groups}
\resizebox{\textwidth}{!}{%
\begin{tabular}{lccc}
\hline
\textbf{Category} & \textbf{Students (n = 538)} & \textbf{TAs (n = 22)} & \textbf{Instructors (n = 6)} \\
\hline
Prerequisite Knowledge & 32\% (n = 172) & 36.4\% (n = 8)\textsuperscript{*} & 16.7\% (n = 1) \\
Time Management & 19.5\% (n = 105) & 22.7\% (n = 5)\textsuperscript{*} & 16.7\% (n = 1) \\
Grade Pressure & 13.6\% (n = 73) & 9.1\% (n = 2) & 33.3\% (n = 2)\textsuperscript{*} \\
Laziness & 11.2\% (n = 60) & 13.6\% (n = 3)\textsuperscript{*} & 33.3\% (n = 2)\textsuperscript{*} \\
Instructor Actions/Perception & 8.4\% (n = 45) & 0\% & 0\% \\
Workload & 7.4\% (n = 40) & 13.6\% (n = 3)\textsuperscript{*} & 0\% \\
No Viable Reason & 2.8\% (n = 15) & 0\% & 0\% \\
Strategy & 1.9\% (n = 10) & 0\% & 0\% \\
Naivety & 1.5\% (n = 8) & 0\% & 0\% \\
Bad Teammates & 1.1\% (n = 6) & 0\% & 0\% \\
Other & 0.7\% (n = 4) & 4.5\% (n = 1) & 0\% \\
\hline
\end{tabular}%
}
\textsuperscript{*}\textit{Denotes a higher percentage than among students; comparisons are descriptive only.}

\label{tab:reasons_comparison}
\end{table}

Our findings show differences in the reasons participants associated with academic dishonesty (see Table \ref{tab:reasons_comparison}). Among the instructors who participated, \textit{Grade Pressure} and \textit{Laziness} were the most frequently mentioned reasons, with two instructors identifying each theme. In contrast, students and TAs most frequently emphasized \textit{Prerequisite Knowledge} and \textit{``Time Management"}. Given the small instructor sample, these findings are descriptive and should not be interpreted as representative of instructors more broadly. 

\textbf{\textit{Prerequisite Knowledge}} was our most frequently identified theme among students (32.0\% of responses, n = 172) and TAs (36.4\% of responses, n = 8). Responses frequently indicated that a lack of adequate prior knowledge or preparation could make succeeding in the course more difficult and could contribute to students relying on forms of assistance they considered cheating. We defined the theme as a lack of necessary skills taught in or before the class that are essential to an assignment or difficulty with the knowledge necessary to complete course work. Below are some examples from students and TAs:

\begin{itemize}
    \item \textit{``Lack of preparation/ familiarity with the material. Time constraint preventing them from properly preparing.''} [S252].

    \item \textit{``It is very difficult and people feel that they are not capable and smart enough to put the effort in to understanding the material.''} [S324].

    \item \textit{``If a student missed class or recitation for a while and does not know what is going on when looking at a programming assignment, they may cheat to figure out stuff they are behind on.''} [S397].

    \item \textit{``I believe one of the reasons that could lead to cheating in this course is lack of understanding on fundamentals needed to accomplish project goals in this class. For example, this class involves the usage of a lot of different technologies the majority of the class has not yet had the opportunity to work with, resulting in students finding the pressure to rely heavily on outside resources.''} [T20].
\end{itemize}

\textbf{\textit{Time Management}} was the second most frequently identified theme among both students and TAs, representing 19.5\% of student responses (n = 105) and 22.7\% of TA responses (n = 5). Participants often indicated that insufficient time, procrastination, or poor planning could lead students to rush or seek shortcuts when completing assignments. We defined it as procrastination; a lack of proper planning and utilization of one's time; or feeling that there was insufficient time to complete an assignment. 

\begin{itemize}
    \item \textit{``Waiting till the last minute to start working on a project and taking shortcuts and cheating to get it finished on time''} [S008].

    \item \textit{``With all of the available resources, I think that poor time management is the main driver of cheating. Especially as a class many first year students take, while they learn to manage the intense course load at [the college] there are often times with too much to do and not enough time to do it. If there is no flexibility with deadlines this can lead to cheating as many students prefer that risk over a 0.''} [S190].

    \item \textit{``Procrastination could definitely be a factor, where if people wait to long on assignments they feel like they need to resort to cheating in order to complete the assignment on time''} [S517].

    \item \textit{``Poor time-management, leading to procrastination and insufficient time to complete the assignment.''} [T23].
\end{itemize}

\textbf{\textit{Grade Pressure}} was mentioned by two of the six instructors who participated and represented the third most frequent theme among students (13.6\% of responses, n = 73). Responses associated with this theme described pressure to maintain or obtain high grades for academic, professional, or personal reasons as a potential motivation for academic misconduct. We defined it as an overwhelming amount of pressure, stress, and desire to get a good grade or meet academic expectations.

\begin{itemize}
    \item \textit{``One obvious reason that would lead a student to use dishonest means in [the class] would be a serious necessity to achieve a certain grade.''} [S058].

    \item \textit{``'I think that someone may end up cheating in [the class] (or in other courses) due to feeling an overwhelming amount of pressure to receive good grades, whether they need to for external reasons (i.e. for scholarships with a required GPA that needs to be maintained) or for internal reasons (i.e. perfectionism).''} [S066].

    \item \textit{``I think that a lot of the freshman coming into this class are getting grades lower than they have before in high school. I think that this might turn many of them to cheating to try and “validate” themselves.''} [T05].

    \item \textit{``At the top of the list, IMO, is the pressure to get high grades to secure internships, grad programs, etc. Overload is another common reason I found among my students. There are many reasons why this could be problematic, including lack of experience managing their agenda, inability to properly assess how much work an individual can take, poor or nonexistent study habits, and/or entitlement due to previous educational experiences.''} [F06].
\end{itemize}

\textbf{\textit{Laziness}} was also mentioned by two of the six instructors who participated. Among students, this theme represented 11.2\% of responses (n = 60), while among TAs it represented 13.6\% of responses (n = 3). Responses within this theme reflected a lack of motivation to learn, complete the work, or spend time on the course. We defined this theme as a lack of desire to learn; impatience; or an unwillingness to spend substantial time on homework or studying.

\begin{itemize}
    \item \textit{``Speed and time. It’s easier to work with someone and be done faster. Plus students often do not care about the definitions of cheating in homework assignments because of the nature of the workplace and collaboration outside of school.''} [S116].

    \item \textit{``The optimal way to get a good grade (whether defined by time spent or work done) is to use all resources available to you. People don't cheat to get a good grade. People cheat to get a good grade faster.''} [S104].

    \item \textit{``It's easy, requires little effort, and often has little repercussions because it's hard to catch.''} [T14].

    \item \textit{``Laziness, lack of passion for the topic''} [F03].
\end{itemize}

\section{Discussion}
This research offers an in-depth look into the perceptions and motivations of cheating through the lens of different actors in the computer science classroom. Our findings indicate that students, TAs, and instructors do not always interpret academic misconduct or its possible motivations in the same way. Within the student population, we also observed small associations between selected demographic characteristics and how some scenarios were classified. These findings reinforce prior work showing that unclear or differently interpreted definitions of cheating can result in varying perceptions of academic dishonesty \cite{3}.

Within the student population, age was negatively associated with cheating classifications in several scenarios. However, these relationships were small in magnitude, and only a subset remained statistically significant after applying the Holm-Bonferroni correction for multiple comparisons. Rather than suggesting that older students generally consider fewer behaviors to be cheating, our findings indicate more limited differences in how students of different ages interpreted particular scenarios. These patterns may be useful for future research examining how students' academic and professional experiences shape their understanding of acceptable collaboration, resource use, and academic misconduct. Our data, however, do not directly establish why these age-related differences occur.

We observed a similar pattern for self-reported religiosity. A small negative association was found between religiosity and the classification of one scenario, with greater self-reported religiosity associated with a slightly lower cheating classification. This association remained small in magnitude and should not be interpreted as evidence of a broader psychological or moral mechanism. In particular, our study did not investigate how participants' religious beliefs influence their moral reasoning or how they reconcile those beliefs with academic-integrity decisions. Future work could examine whether and how religiosity relates to interpretations of academic misconduct using measures specifically designed to investigate these processes.

We also observed differences in the reasons participants associated with academic dishonesty. Among the six instructors who participated, \textit{Grade Pressure} and \textit{Laziness} were the most frequently identified themes, whereas students and TAs more frequently emphasized \textit{Prerequisite Knowledge} and \textit{Time Management.} Given the small instructor sample, these differences should be interpreted descriptively rather than as representative differences between faculty and students more broadly. Nevertheless, the pattern is consistent with previous work emphasizing the importance of communication, classroom support, and clear expectations in shaping academic-integrity experiences \cite{9,13}.

The open-ended responses further illustrate how participants may understand responsibility for academic difficulties in different ways. For example, a lack of prerequisite knowledge can place greater emphasis on students' preparation for a course or on the instructional support available to them, whereas laziness places greater emphasis on individual student motivation and behavior. In several responses, participants explicitly connected academic misconduct to perceived shortcomings elsewhere in the classroom environment. On the student side, we find responses like \textit{``Due to a lack of proper instructional time and in-depth discussions on the course contents that actually matter, there are many unaddressed nuances and confusions related to the course that would require asking a lot of questions to TA's and possibly cheating to understand the content properly."} [S007], or sometimes simply \textit{``The professors aren't explaining well"} [S156]. Conversely, on the instructor side we find \textit{``Laziness; Lack of passion for the topic; Inability to manage time properly; Lack of knowledge, skill, or ability; Different cultural attitudes to cheating; Lack of motivation; Lack of moral character"} [F03]. These responses suggest that students and instructors may sometimes locate the causes of academic misconduct in different parts of the educational experience. Rather than interpreting this as blame shifting, we see it as further evidence of the importance of creating opportunities for students and instructional staff to communicate their expectations, challenges, and responsibilities more explicitly.

Overall, our study underscores the importance of communication among students, TAs, and instructors regarding academic misconduct, particularly because the same scenarios and possible motivations can be understood differently by different actors in the classroom. This aligns with previous research emphasizing that clear expectations and explicit boundaries can reduce ambiguity surrounding academic integrity \cite{3,13}. Our findings extend this discussion by showing how these differences appear across students, TAs, and instructors responding to the same set of scenarios in contemporary CS courses.

Our demographic findings also provide directions for future research, but they should be interpreted as secondary and exploratory. The small associations involving age and self-reported religiosity suggest that individual characteristics may relate to how some academic-integrity situations are interpreted, but our study does not explain the mechanisms behind these relationships. Future work could examine these associations using larger and more diverse samples and measures specifically designed to investigate why such differences emerge. More broadly, understanding where perceptions differ across classroom actors represents an important step toward developing clearer expectations and a more shared understanding of academic integrity in computing education.

\section{Limitations}
Some of the main limitations of this study include its single-institution setting, the demographics and academic level of the participating courses, the unequal number of participants across stakeholder groups, and the scenario-based and self-reported nature of our data. This study was conducted at a large southeastern university in the United States, which may not reflect the CS educational space holistically. Although we surveyed a large number of students, the size of the student sample does not eliminate the limitations associated with recruiting participants from a single institution and two courses. Therefore, the generalizability of our findings to other institutions, programs, and student populations may be limited.

Similarly, as this study was conducted in two large-enrollment courses ranging from introductory programming to introductory software design and engineering, the student demographics heavily favored younger students and underclassmen. We would like to conduct follow-up studies with more advanced computing classes to examine whether the patterns observed in this study also appear among populations with greater representation from junior and senior students.

The number of participants was also substantially different across the three stakeholder groups, with 538 students, 22 TAs, and six instructors. This is particularly important for the faculty group. With only six participating instructors, the findings from this group should be considered descriptive and exploratory rather than representative of computing faculty more broadly. The unequal sample sizes also limit the extent to which comparisons among students, TAs, and instructors can be interpreted as broader group-level differences. Future work should recruit larger and more balanced samples of instructional staff.

Our recruitment process introduces additional limitations. Students received extra credit for participating, whereas TAs and faculty were not offered an incentive. Additionally, because we were unable to reliably determine the total number of individuals who received the study invitation, we could not calculate response rates. As a result, self-selection and non-response bias may have influenced which participants chose to complete the study.

The study also relies on participants' self-reported perceptions of hypothetical academic-integrity scenarios. How participants classify a scenario does not necessarily indicate how they would behave when facing a similar situation in an actual course. Likewise, the scenarios necessarily simplify situations that may depend on course policies, assignment instructions, instructor expectations, and other contextual factors. While the scenarios allowed us to compare responses to a common set of situations, future research could complement this approach with interviews, observations, or analyses of actual academic-integrity cases.

Finally, the demographic associations identified in the student population were small in magnitude. Although we applied the Holm-Bonferroni correction to account for multiple comparisons, these findings remain exploratory and should not be interpreted as evidence that characteristics such as age or self-reported religiosity broadly determine how students understand academic misconduct. Further work with more diverse samples and measures designed specifically to examine these relationships would help clarify the patterns observed here. 

\section{Conclusions}
Based on our findings and results, there is considerable diversity in how potential academic-integrity violations are interpreted across students, TAs, and instructors, including scenarios involving the use of AI in CS courses. This indicates a need for clearer communication of academic-integrity expectations in the classroom: when students, TAs, and instructors develop differing views on what constitutes cheating, it becomes difficult for the broader student population to determine what is acceptable. Our open-ended responses also showed differences in the reasons participants associated with academic misconduct. Among the instructors who participated, \textit{Grade Pressure} and \textit{Laziness} were the most frequently identified themes, while students and TAs more frequently emphasized \textit{Prerequisite Knowledge} and \textit{Time Management.} Given the small instructor sample, these findings should be considered descriptive rather than representative of instructors more broadly.

We also found small associations between selected student characteristics and how particular scenarios were classified. Age was negatively associated with cheating classifications in several scenarios, although only a subset of these associations remained statistically significant after the Holm-Bonferroni correction. Self-reported religiosity was also negatively associated with the classification of one scenario. These relationships were small in magnitude and do not indicate that age or religiosity broadly determine how students understand academic misconduct. Rather, they provide exploratory evidence that individual characteristics may relate to the interpretation of particular academic-integrity situations and warrant further investigation.

Together, our findings highlight the need for clearer and more explicit boundaries around acceptable AI use and collaboration in computing courses. Instructors often encourage students to treat AI tools as they would a classmate. Nonetheless, our observations suggest that the boundaries surrounding peer collaboration can themselves be interpreted differently across classroom actors. As a result, framing AI use through this analogy may not be sufficient for students to interpret academic integrity in practice.

Overall, our work suggests the need for more explicit communication about acceptable collaboration practices and the role of AI-assisted work in computing education. As AI-assisted programming and other forms of digital collaboration become increasingly common in both educational and professional environments, further research is needed to understand how academic-integrity policies can evolve while maintaining clear expectations and supporting learning and assessment goals. More broadly, developing a shared understanding of what constitutes acceptable assistance may help students, TAs, and instructors navigate academic integrity more consistently in contemporary computing courses.

\section*{Conflict of Interest Statement}
The authors declare that the research was conducted in the absence of any commercial or financial relationships that could be construed as a potential conflict of interest.

\section*{Funding}
This material is based upon work supported by the National Science Foundation under Grant No. 2434428. Any opinions, findings, and conclusions or recommendations expressed in this material are those of the author(s) and do not necessarily reflect the views of the National Science Foundation.

\section*{Acknowledgments}
The authors acknowledge the use of Grammarly and ChatGPT-5.3, an AI-based language model developed by OpenAI, for spell-checking, grammar, and editing assistance. Both tools were used to refine the language and enhance the manuscript's clarity. No content, figures, or images were generated by AI for this work.

\bibliographystyle{Frontiers-Harvard} 
\bibliography{bibliography}

@misc{thematic,
  author       = {Maria Rosala},
  title        = {Analyze Qualitative Data: Thematic Analysis},
  year         = {2022},
  month        = aug,
  publisher    = {Nielsen Norman Group},
  url          = {https://www.nngroup.com/articles/thematic-analysis/},
  note         = {Accessed: 2026-05-18}
}

@inproceedings{1,
  title={Examinator v3. 0: cheating detection in online take-home exams},
  author={Hung, Jui-Tse and Cui, Christopher and Agarwal, Varun and Chatterjee, Saurabh and Apoorv, Raghav and Graziano, Rocko and Starner, Thad},
  booktitle={Proceedings of the Tenth ACM Conference on Learning@ Scale},
  pages={401--405},
  year={2023}
}

@inproceedings{2,
  title={Answer Watermarking: Using Answer Generation Assistance Tools to Find Evidence of Cheating},
  author={Cui, Christopher and Hung, Jui-Tse and Sharma, Pranav and Chatterjee, Saurabh and Starner, Thad},
  booktitle={Proceedings of the Eleventh ACM Conference on Learning@ Scale},
  pages={519--523},
  year={2024}
}

@inproceedings{3,
  title={Causes for cheating: Unclear expectations in the classroom},
  author={Ozment, James A and Smith, Alison N and Newstetter, Wendy},
  booktitle={2000 Annual Conference},
  pages={5--139},
  year={2000}
}

@inproceedings{4,
  title={Examinator v4. 0: Cheating Detection in Online Take-Home Exams},
  author={Cui, Christopher and Hung, Jui-Tse and Malhotra, Vaibhav and Goel, Hardik and Apoorv, Raghav and Starner, Thad},
  booktitle={Proceedings of the Eleventh ACM Conference on Learning@ Scale},
  pages={330--334},
  year={2024}
}

@INPROCEEDINGS{5,
  author={Salhofer, Peter},
  booktitle={2017 IEEE Global Engineering Education Conference (EDUCON)}, 
  title={Analysing student behavior in CS courses: A case-study on detecting and preventing cheating}, 
  year={2017},
  volume={},
  number={},
  pages={1426-1431},
  doi={10.1109/EDUCON.2017.7943035}}

@article{6,
  title={Scaling anti-plagiarism efforts to meet the needs of large online computer science classes: Challenges, solutions, and recommendations},
  author={Adkins, Keith L and Joyner, David A},
  journal={Journal of Computer Assisted Learning},
  volume={38},
  number={6},
  pages={1603--1619},
  year={2022},
  publisher={Wiley Online Library}
}

@article{7,
  title={Plagiarism in programming assessments: a systematic review},
  author={Albluwi, Ibrahim},
  journal={ACM Transactions on Computing Education (TOCE)},
  volume={20},
  number={1},
  pages={1--28},
  year={2019},
  publisher={ACM New York, NY, USA}
}

@inproceedings{8,
author = {Denzler, Benjamin and Vahid, Frank and Pang, Ashley},
title = {Style Anomalies Can Suggest Cheating in CS1 Programs},
year = {2024},
isbn = {9798400704246},
publisher = {Association for Computing Machinery},
address = {New York, NY, USA},
url = {https://doi.org/10.1145/3626253.3635519},
doi = {10.1145/3626253.3635519},
booktitle = {Proceedings of the 55th ACM Technical Symposium on Computer Science Education V. 2},
pages = {1624–1625},
numpages = {2},
location = {Portland, OR, USA},
series = {SIGCSE 2024}
}

@inproceedings{9,
author = {Wortzman, Brett and Stephens-Martinez, Kristin and Minnes, Mia and Ola, Oluwakemi and Blank, Adam},
title = {Who's Cheating Whom: Changing the Narrative Around Academic Misconduct},
year = {2023},
isbn = {9781450394338},
publisher = {Association for Computing Machinery},
address = {New York, NY, USA},
url = {https://doi.org/10.1145/3545947.3569609},
doi = {10.1145/3545947.3569609},
booktitle = {Proceedings of the 54th ACM Technical Symposium on Computer Science Education V. 2},
pages = {1210–1211},
numpages = {2},
location = {Toronto ON, Canada},
series = {SIGCSE 2023}
}

@inproceedings{payne2025exploring,
  title={Exploring Community Perceptions and Experiences Towards Academic Dishonesty in Computing Education},
  author={Payne, Chandler C and Hackney, Kai A and Zangari, Lucas Guarenti and Munoz, Emmanuel and Kalogeras, Sterling R and S{\'a}nchez-G{\'o}mez, Juan Sebasti{\'a}n and Omojokun, Olufisayo and Feij{\'o}o-Garc{\'\i}a, Pedro Guillermo},
  booktitle={Proceedings of the 2025 ACM Conference on International Computing Education Research V. 2},
  pages={13--14},
  year={2025}
}

@inproceedings{10,
author = {Vahid, Frank and Downey, Kelly and Pang, Ashley and Gordon, Chelsea},
title = {Impact of Several Low-Effort Cheating-Reduction Methods in a CS1 Class},
year = {2023},
isbn = {9781450394314},
publisher = {Association for Computing Machinery},
address = {New York, NY, USA},
url = {https://doi.org/10.1145/3545945.3569731},
doi = {10.1145/3545945.3569731},
booktitle = {Proceedings of the 54th ACM Technical Symposium on Computer Science Education V. 1},
pages = {486–492},
numpages = {7},
location = {Toronto ON, Canada},
series = {SIGCSE 2023}
}

@inproceedings{11,
author = {Xiao, Ruiwei and Huerta-Mercado, Eduardo and Garcia, Dan},
title = {Detecting Cheating in Online Take-Home Exams with Randomized Questions},
year = {2023},
isbn = {9781450394338},
publisher = {Association for Computing Machinery},
address = {New York, NY, USA},
url = {https://doi.org/10.1145/3545947.3576270},
doi = {10.1145/3545947.3576270},
booktitle = {Proceedings of the 54th ACM Technical Symposium on Computer Science Education V. 2},
pages = {1323},
numpages = {1},
location = {Toronto ON, Canada},
series = {SIGCSE 2023}
}

@inproceedings{12,
author = {Vykopal, Jan and \v{S}v\'{a}bensk\'{y}, Valdemar and Seda, Pavel and \v{C}eleda, Pavel},
title = {Preventing Cheating in Hands-on Lab Assignments},
year = {2022},
isbn = {9781450390705},
publisher = {Association for Computing Machinery},
address = {New York, NY, USA},
url = {https://doi.org/10.1145/3478431.3499420},
doi = {10.1145/3478431.3499420},
booktitle = {Proceedings of the 53rd ACM Technical Symposium on Computer Science Education - Volume 1},
pages = {78–84},
numpages = {7},
location = {Providence, RI, USA},
series = {SIGCSE 2022}
}

@inproceedings{13,
author = {Mason, Tony and Gavrilovska, Ada and Joyner, David A.},
title = {Collaboration Versus Cheating: Reducing Code Plagiarism in an Online MS Computer Science Program},
year = {2019},
isbn = {9781450358903},
publisher = {Association for Computing Machinery},
address = {New York, NY, USA},
url = {https://doi.org/10.1145/3287324.3287443},
doi = {10.1145/3287324.3287443},
booktitle = {Proceedings of the 50th ACM Technical Symposium on Computer Science Education},
pages = {1004–1010},
numpages = {7},
location = {Minneapolis, MN, USA},
series = {SIGCSE '19}
}

@inproceedings{14,
author = {Pang, Ashley and Vahid, Frank},
title = {ChatGPT and Cheat Detection in CS1 Using a Program Autograding System},
year = {2024},
isbn = {9798400706004},
publisher = {Association for Computing Machinery},
address = {New York, NY, USA},
url = {https://doi.org/10.1145/3649217.3653558},
doi = {10.1145/3649217.3653558},
booktitle = {Proceedings of the 2024 on Innovation and Technology in Computer Science Education V. 1},
pages = {367–373},
numpages = {7},
location = {Milan, Italy},
series = {ITiCSE 2024}
}

@inproceedings{15,
author = {Savelka, Jaromir and Agarwal, Arav and Bogart, Christopher and Song, Yifan and Sakr, Majd},
title = {Can Generative Pre-trained Transformers (GPT) Pass Assessments in Higher Education Programming Courses?},
year = {2023},
isbn = {9798400701382},
publisher = {Association for Computing Machinery},
address = {New York, NY, USA},
url = {https://doi.org/10.1145/3587102.3588792},
doi = {10.1145/3587102.3588792},
booktitle = {Proceedings of the 2023 Conference on Innovation and Technology in Computer Science Education V. 1},
pages = {117–123},
numpages = {7},
location = {Turku, Finland},
series = {ITiCSE 2023}
}

\end{document}